\documentclass[twocolumn,superscriptaddress,
showpacs,preprintnumbers,amsmath,amssymb]{revtex4}

\usepackage{graphicx}
\usepackage{dcolumn}
\usepackage{bm}
\usepackage{epsfig}

\newcommand{\bra}[1]{{\left\langle #1 \right|}}
\newcommand{\ket}[1]{{\left| #1 \right\rangle}}
\newcommand{\cA}{{\cal A}}
\newcommand{\cM}{{\cal M}}
\newcommand{\cS}{{\cal S}}

\newcommand{\bH}{{\mathbb H}}
\newcommand{\tr}{{\mathop{\mathrm{tr}}}}

\begin{document}


\title{Quantum Solution to the Extended Newcomb's Paradox}

\author{Dong Pyo Chi}\email{dpchi@math.snu.ac.kr}
\affiliation{
 School of Mathematical Sciences,
 Seoul National University, Seoul 151-742, Korea
}
\author{Kabgyun Jeong}\email{kgjeong6@snu.ac.kr}
\affiliation{ Interdisciplinary Program in Nano-Science and
Technology, Seoul National University,  Seoul 151-742, Korea
}

\date{\today}

\begin{abstract}

We regard the Newcomb's Paradox as a reduction of the Prisoner's
Dilemma and search for the considerable quantum solution. The all
known classical solutions to the Newcomb's problem always imply
that human has freewill and is due to the unfair set-up(including
strategies) of the Newcomb's Problem. For this reason, we here
substitute the asymmetric payoff matrix to the general form of the
payoff matrix~($\cM$) and consider both of them use the same
quantum strategy. As a result we obtained the fair Nash
equilibrium, which is better than the case using classical
strategies. This means that whether the supernatural being has the
precognition or not depends only on the choice of strategy.

\end{abstract}


\maketitle

\section{\label{sec:level1}Introduction}

A quantum game theory is a natural extension to the quantum world
of the ubiquitous classical game theory. Von~Neumann pointed
out~\cite{NM}, a game is simply the totality of the rule which
describe it. The elements of the rule such as the rational
choices(strategies) are one of the main component of a general
game. Moveover a game assumed that each individual player is
trying to maximize his own advantage, without concern for the
well-being of the other players. After the theory of the ganeral
game has been broken down into details~\cite{M}, we can meet
several games to play more efficiently for his profits. For
example, von~Neumann's the mathematical approaches help us
rational decision for the economic behavior, which strategies are
fit and what's the causal relationship. The Darwin's theory for
the \emph{'survival of fittest'} is maybe the another well-known
game, the evolutionary theory, describe the biological evolution
in nature, it is described by the Evolutionarily Stable
Strategy(ESS)~\cite{IT}.

Here, we present two games known as the Prisoner's
dilemma(PD)~\cite{EWM} and the Newcomb's Paradox(NP)~\cite{PS}.
Those game have some irrational consequence, {\it i.e.,} the
players cannot concurrently increase their payoff using a certain
strategy in the PD. Also in the NP case, the supernatural
being($\cal {SB}$)'s prophecy fail to facing human's freewill
surprisingly. Game theorists maybe call this ambiguous situation
as a dilemma or paradox for a given competition. We will solve the
NP under the well-known PD solution, which means the PD's
symmetric consideration helpful to grasp the NP.

The classical games, as long as its history, form a various kinds
of game contained certain rules and it also have studied deeply.
Its solutions of the dilemmatic view point are generally solved by
the Nash equilibrium governed by a dominant strategy. While the
usual quantum games~\cite{EWM,Meyer} have comparatively short
history, but their solution shows the similar Nash equilibrium
pattern--- the changed Nash equilibrium can be compared to the
classical one. Meyer has already proved~\cite{Meyer} that an
optimal quantum strategy in a two-person zero-sum game surpass
against to an optimal classical~(mixed) strategy. If the
participant, Alice, is using a probabilistic~(mixed) quantum
strategies against to Bob's the deterministic classical
strategies, then Alice always win to Bob on the game. Furthermore
a two-person static game has a Nash equilibrium for the quantum
versus quantum strategy. For consider the NP, $\cal {SB}$'s
deterministic prophecy gives an unfair situation for the human's
probabilistic strategies. So we need to reconstruct the original
payoff to the new reasonable payoff structure.

\section{\label{sec:leve21}Classical Strategy and Its Solution}
\subsection{\label{sec:leve21}NP and PD in the Classical Game}

In the PD each of the players, Alice and Bob, must independently
decide whether they choose to defect~(strategy $D$) or
cooperate~(strategy $C$). Depending on their decision, each player
receives a certain payoff $\Pi_k$. The strategy $D$ is the
dominant and will be in equilibrium, {\it i.e.,} $\{\Sigma_A,
\Sigma_B\}=\{D, D\}$ and
$\{\Pi_A(\Sigma_A),\Pi_B(\Sigma_B)\}=\{1,1\}$ in this game. A
dominant strategy for Alice is defined by a strategy $\Sigma_A$
such that the payoff $\Pi_A$ has the property  $\Pi_A(\Sigma_A,
{\Sigma^j}_B)\geq \Pi_A({\Sigma^i}_A,{\Sigma^j}_B)$ for all
${\Sigma^i}_A\in{S_A},{\Sigma^j}_B\in{S_B}$ provided such a
strategy exists, where $i, j\in N$. The strategy set
$S_k\simeq\{{\Sigma^i}_k\}$ contains the all strategies $i\in N$,
where $N$ is the number of the all possible strategies in the game
and $k$ denote the players $A$ or $B$. Moreover Nash equilibrium,
corresponding to the payoff $\{1, 1\}$, is a combination of
strategies $\{\Sigma_A, \Sigma_B\}$ such that neither party can
increase his or her payoff by unilaterally departing from the
given equilibrium point {\it i.e.,} $\Pi_A(\Sigma_A, \Sigma_B)\geq
\Pi_A({\Sigma^i}_A,\Sigma_B)$ and $\Pi_B(\Sigma_A, \Sigma_B)\geq
\Pi_B(\Sigma_A,{\Sigma^j}_B)$. Another main concept of a game
theory is the {\it Pareto} optimality, a pair of payoffs
$\{\Sigma_A, \Sigma_B\} = \{3, 3\}$, if it is not jointly
dominated by another point, and if neither party can increase his
or her payoff without decreasing the payoff to the other party.
Someone perhaps ask that the strategy $\{C, C\}$ is more rational
decision. But the dominant strategy $D$ controls the PD.

The NP is said to be a one-person asymmetric game such that the
human's freewill conflict against to the $\cal {SB}$'s prophecy.
That is, given two boxes, $Box1$ and $Box2$, $Box2$ which contains
${\$1~000}$ and $Box1$ which contains either nothing or
${\$1~000~000}$, the human may pick either $Box1$ or both.
However, at some time before the choice is made by human, $\cal
{SB}$ has predicted what the human's decision will be
made~\cite{Jeong}. Thereafter $\cal {SB}$ will fill up the $Box1$
with ${\$1~000~000}$ if $\cal {SB}$ predicts human to take it, or
with nothing if $\cal {SB}$ predicts human to take both boxes. The
constructionally important notion of the NP is no other than its
one-person game as the restricted two-person game PD's~\cite{Sch}.
We can see the original asymmetric payoff matrix for human in
Table~\ref{tab:table2}.

\begin{table}[tbp]
\caption{\label{tab:table1}The payoff matrix for the traditional
Prisoner's Dilemma. The first entry in the round bracket denotes
the payoff of Alice and the second one the payoff of Bob.}
\begin{ruledtabular}
\begin{tabular}{ccc}
           &Bob: $C$  & Bob: $D$\\
\hline
Alice: $C$ & $(3, ~3)$ & $(0, ~5)$\\
Alice: $D$ & $(5, ~0)$ & $(1, ~1)$\\
\end{tabular}
\end{ruledtabular}
\end{table}

\begin{table}[tbp]
\caption{\label{tab:table2}The payoff matrix for the original
Newcomb's Paradox. The elements of the matrix will give to human
for his reward.}
\begin{ruledtabular}
\begin{tabular}{ccc}
&$\cal {SB}$: $Box1$&$\cal {SB}$: $Box2$\\
\hline
Human: $Box1$ & ${\$1~000~000}$ & ${0}$\\
Human: $Box2$ & ${\$1~001~000}$ & ${\$1~000}$\\
\end{tabular}
\end{ruledtabular}
\end{table}

\subsection{\label{sec:leve22}Extended form of NP and PD}
Von~Neumann referred to the fact that game theory only concern
with the relation of the elements of the payoff matrix. So that in
this paper, we can consider the extended version of the NP and PD.
The general form of the corresponding payoff matrix is denoted by
$\cM=\cM_1+\cM_2$
where $\cM_1 = \left(%
\begin{array}{cc}
  \alpha & 0 \\
  0 & \gamma \\
\end{array}%
\right)$ and $\cM_2 = \left(%
\begin{array}{cc}
  0 & \delta \\
  \beta & 0 \\
\end{array}%
\right)$, $\cM_2$ will be considered as two special case, an
asymmetric payoff matrix and the quasi-skew symmetric matrix.
Moreover we fix the matrix elements, in the payoff matrix, along
to $\beta>\alpha>\gamma>\delta$ and $\delta\leq0$.\\

{\it (i) Asymmetric payoff matrix $\cA$:} If the matrix $\cM$ is
satisfied some next property
\begin{eqnarray}\cA = \cA_1 + \cA_2\end{eqnarray}
where $\cA_1=\cM_1$ is a diagonal
matrix, $\cA_2=\left(%
\begin{array}{cc}
  0 & \delta \\
  \beta & 0 \\
\end{array}%
\right)\neq{\cA_2}^T$ is an off-diagonal asymmetric matrix with
$|\delta|<\beta$, then we call the matrix $\cA$ as the asymmetric.
Specially, the case of $\delta=0$ occurs the original NP and PD.\\

{\it (ii) Quasi-skew symmetric payoff matrix $\cS$:} The payoff
matrix $\cM$ is defined as a quasi-skew symmetric matrix $\cS$
{\it i.e.,} decomposed by
\begin{eqnarray}
\cS = \cS_1 + \cS_2
\end{eqnarray}
where $\cS_1 = \cM_1$ is a diagonal matrix, $\cS_2 = \left(%
\begin{array}{cc}
  0 & -\beta \\
  \beta & 0 \\
\end{array}%
\right)$ is the skew-symmetric matrix which satisfies the property
${\cS_2}^T = -{\cS_2}$. If we generally allow the extended payoff
matrix $\cM$ to the traditional NP, then one can naturally extend
the matrix as corresponding two-person symmetric PD in
Table~\ref{tab:table3}-\ref{tab:table4}. As we see previously, the
payoff matrix of the symmetric PD is more clear to interpret and
easy to calculate the solution. Thereafter we will change the
formation from the two-person's PD to the one-person's NP. Here we
must consider a sense of the leaping element $\delta$ in the
matrix $\cM$. The payoff values, all elements of the matrix for
human in the original NP, have always a positive value more than
$0$. This means that the payoffs are only beneficial to human,
moreover he does cheat $\cal SB$ using the convex combination of
the classical mixed strategies. The set-up is unreasonable and
one-sided {\it i.e.,} trying wouldn't do any harm for human. If we
give a negative payoff element $-\beta$, the case {\it (ii)}, for
the player such as $\cal SB$'s penalty for human's cheating or
human's entry fee, perhaps some anticipointment for an untruth of
$\cal SB$'s prophecy. Likewise the solution for the PD, the
element $\delta\neq-\beta$ also gives a corresponding Nash
equilibrium. We adapt the Nash equilibrium consider as the quantum
solution to the NP {\it i.e.,} if human choose $\alpha$ then $\cal
SB$ can exactly predict the human's choice $\alpha$.
\begin{table}[tbp]
\caption{\label{tab:table3}The payoff matrix for the extended
Newcomb's Paradox. The entry $\delta(<0)$ restrict the human's
choice.}
\begin{ruledtabular}
\begin{tabular}{ccc}
&$\cal {SB}$: $Box1$&$\cal {SB}$: $Box2$\\
\hline
Human: $Box1$ & ${\alpha}$ & ${\delta}$\\
Human: $Box2$ & ${\beta}$  & ${\gamma}$\\
\end{tabular}
\end{ruledtabular}
\end{table}

\begin{table}[tbp]
\caption{\label{tab:table4}The payoff matrix for the extended
Prisoner's Dilemma. Alice correspond to human and Bob to $\cal
SB$, also $C$ change to the $Box1$ strategy and $D$ equal to the
$Box2$.}
\begin{ruledtabular}
\begin{tabular}{ccc}
           &$\cal SB$: $Box1$  & $\cal SB$: $Box2$\\
\hline
Human: $Box1$ & $(\alpha,~\alpha)$ & $(\delta,~\beta)$\\
Human: $Box2$ & $(\beta,~\delta)$  & $(\gamma,~\gamma)$\\
\end{tabular}
\end{ruledtabular}
\end{table}

\section{\label{sec:leve31}Quantum Strategy}
\subsection{\label{sec:leve31}Quantum Strategies and Quantum Circuit}

A general quantum game $\Gamma$ is consisted of the player $k$,
strategy $\hat\Sigma_k\in \hat S_k\subset\hat U_k\equiv\hat
U_k(\theta,\phi)$ and payoff $\Pi_k$ like as the general classical
game~\cite{G}. But the strategic space $\hat S_k$ is comparatively
large space chosen from $\hat U$, $\hat S_k\subset\hat U_k
\subseteq\hat U$, $\hat U\equiv\{\hat
U(\theta,\phi)|\theta\in[0,\pi]$ and $\phi \in[0, {\pi}/2]\}$ in
Hilbert space $\bH$, compare to the classical deterministic space
$S_k$. The quantum games are following the next trivial stages.
First, the well-known initial states are prepared by a umpire or
the public renouncement, because the fairness of the game is
ensured by its open source of the payoff. In PD, we set the
initial state as two qubit state $\ket{00}$. Next step, both
player conflict to win the game with the given strategy set $\hat
S_k\subset\hat U_k$, which is the quantum mechanical operation.
Main point of a game maybe make out what is the best strategy in
the confliction. Finally, the last step is some detection of the
game result by quantum measurement.

We denote the initial state of the game by $\ket{\psi_0} =
\hat{J}\ket{00}$, where $\hat{J}$ is a unitary operation executed
by an umpire and the adjustment of the parameter $\tau$ gives a
separable or entangled states. The unitary operation $\hat{J}$ is
described by $\hat{J}= e^{\{i\tau {\tilde U\otimes\tilde U}/2\}}$,
where $\tau\in[0, \pi/2]$ and $\tilde U\in \hat U(\theta, \phi)$
stand for the selected operation by an umpire. In fact, $\tau$ is
said to be a measure for the game's entanglement {\it i.e.,}
$\tau=\pi/2$ means the maximally entangled state. If we have
prepared the initial game setup such as the separable or entangled
state, then we can paly PD in each player's strategic space $\hat
S_k$. Each player may execute own's qubit by $\hat\Sigma_k\in \hat
S_k$ for increment of his payoff, of course the player let to
moves secretly. It provide to be sufficient to restrict the
strategic space to the two-parameter set of $2\times2$ matrices,
\begin{eqnarray}
\hat U(\theta,\phi) = \left(%
\begin{array}{cc}
  e^{i\phi}\cos{\theta/2} & \sin{\theta/2} \\
  -\sin{\theta/2} & e^{-i\phi}\cos{\theta/2} \\
\end{array}%
\right)
\end{eqnarray}
where $0\leq\theta\leq\pi$ and $0\leq\phi\leq\pi/2$. From the
result of the Eisert {\it et al.}~\cite{EWM,BH}, where various
kinds of strategy are introduced, we can find the general quantum
leaps in the game theory. Specially the strategy $\hat
Q\otimes\hat Q\equiv\hat U(0, \pi/2)\otimes\hat U(0, \pi/2)$ give
a new Nash equilibrium, against to classical strategy $\hat
D\otimes\hat D \equiv\hat U(\pi, 0)\otimes\hat U(\pi, 0)$. Here,
we will briefly depict the evolution of qubit state. The prepared
initial state of the two qubit, to play on the game, is denoted by
$\ket{\psi_0}=\hat J\ket{00}$. Next the participants, Alice and
Bob, play own's qubit using the unitary operation $\hat U_k$. The
state of the game is prepared, after the previous strategic
process, like that
\begin{eqnarray}
\ket{\psi_f}={\hat J}^\dagger(\hat U_A \otimes \hat U_B)\ket
{\psi_0}
\end{eqnarray}
where $\ket{\psi_f}$ means the final state of the game, see figure
in \cite{EWM}. Before we measure the final state, the Alice's
payoff can be expected as $\bar{\Pi}_A = \alpha P_{\hat C\hat C} +
\beta P_{\hat D\hat C} + \delta P_{\hat C\hat D} + \gamma P_{\hat
D\hat D}$ where
$P_{\hat\Sigma_A\hat\Sigma_B}=|\langle{\hat\Sigma_A\hat\Sigma_B}|
{\psi_f}\rangle|^2$ is the measurable probability with
corresponding payoff as the coefficient respectively. If the each
qubit is measured by appropriate device, then each player obtains
the corresponding payoff which is a resultant value of the quantum
PD. For example, the case of $\tau = \pi/2$ and $\hat Q \otimes
\hat Q$ build up to the new Nash equilibrium $\{\alpha, \alpha\}$
from the classical equilibrium $\{\gamma, \gamma\}$. That is, the
classical {\it Pareto} optimal correspond to the new stable point
of the PD.

Now we will compute the new Nash equilibrium in the PD, which is
occurred by the some additional quantum strategy such as
Walsh-Hadamard transformation $H$,
\begin{eqnarray}
H = \frac{1}{\sqrt{2}}\left(%
\begin{array}{cc}
  1 & 1 \\
  1 & -1 \\
\end{array}%
\right)
\end{eqnarray}
or Pauli operations
\begin{eqnarray}
\sigma_x = \left(%
\begin{array}{cc}
  0 & 1 \\
  1 & 0 \\
\end{array}%
\right),~
\sigma_y = \left(%
\begin{array}{cc}
  0 & -i \\
  i & 0 \\
\end{array}%
\right),~
\sigma_z = \left(%
\begin{array}{cc}
  1 & 0 \\
  0 & -1 \\
\end{array}%
\right).
\end{eqnarray}
First of all the initial state $\ket {\psi_0}$ let to be the
maximally entangled state, changing the variational value $\tau$
from $0$ to $\pi/2$ and the choice of $\tilde U$. For instance, if
the entangled measure $\tau$ fix to $\pi/2$ and the umpire's
choice $\tilde U$ select to $\hat D\otimes\hat D$, then the value
$\hat J$ is locked up~\cite{EWM,G} like that $\hat J =
e^{(i\pi{\hat D\otimes\hat D}/4)} =\frac{1}{\sqrt2}({\hat
I}^{\otimes2}+i{\hat D}^{\otimes2})$, where $\hat I$ means the
identity operation. Therefore, we obtain the maximally entangled
state, denoted by $\tilde{\ket{\psi_0}} = \hat J\ket{00} =
\frac{1}{\sqrt2}(\ket{00}+i\ket{11})$, and sequentially get to the
separable final state $\tilde{\ket{\psi_f}}$, by the disentangling
operation ${\hat J}^\dagger$ after the local operations $\tilde
U_A = \hat D$ and $\tilde U_B = \hat D$ are done of each player
respectively {\it i.e.,} $\tilde{\ket{\psi_f}}
 = \frac{1}{\sqrt2}{\hat J}^\dagger(\ket{11}+i\ket{00})
 = \ket{11}$.
Although we perform the measurement by a certain observable
$\hat\Omega_{A(B)}$ with the combinations of
\begin{eqnarray}
\{\alpha\ket{00}\bra{00}, \delta(\beta)\ket{01}\bra{01},
\beta(\delta)\ket{10}\bra{10}, \gamma\ket{11}\bra{11}\},
\end{eqnarray}
the Nash equilibrium,
\begin{eqnarray}
\{\bar\Pi_A&=&\tr({\hat\Omega_A{\tilde{\ket{\psi_f}}\tilde{\bra{\psi_f}}}}),\nonumber\\&&
{\bar\Pi_B=\tr({\hat\Omega_B{\tilde{\ket{\psi_f}}\tilde{\bra{\psi_f}}}}})\}
= \{\gamma, \gamma\}
\end{eqnarray}
with the probability $1$, is obtained on this traditional
strategy. This is only the classical consequence.

But if we choose the some less quantum strategy and full quantum
strategies such as the Hadamard transformation $H$ and Pauli
matrices $\{\sigma_x, \sigma_y, \sigma_z\}$, then the new Nash
equilibrium is created, that is, we will choose the extended
strategic space $\hat S$. Given the entangled state
\begin{eqnarray}
\tilde{\ket{\psi_0}}=\frac{1}{\sqrt2}(\ket{00}-i\ket{11})
\end{eqnarray}
with $\hat J=e^{i\pi(\sigma_y\otimes\sigma_y)/4}$ transform to the
final state via the local operation ${\tilde U}_k=H_k\in\hat
U_k(\theta, \phi)$,
\begin{eqnarray}
\tilde{\ket{\psi_f}}
 &= \frac{1}{2}(\ket{00}-\ket{01}-\ket{10}+\ket{11}).
\label{eq:state}
\end{eqnarray}
The final state~(\ref{eq:state}) gives
$\bar\Pi_A=\tr({\hat\Omega_A{\tilde{\ket{\psi_f}}\tilde{\bra{\psi_f}}}})
=\frac{1}{4}(\alpha+\delta+\beta+\gamma)$, that is,
\begin{eqnarray}\{\bar\Pi_A,\bar\Pi_B\}=\{{\frac{1}{4}(\alpha+
\delta+\beta+\gamma),\frac{1}{4}(\alpha+\beta+\delta+\gamma)}\}.
\label{eq:equilib}
\end{eqnarray}
Equation~(\ref{eq:equilib}) is always superior to the classical
equilibrium $\{{\frac{1}{4}(\alpha+
\gamma),~\frac{1}{4}(\alpha+\gamma)}\}$. But the value $\delta$
equal to $-\beta$, the skew-symmetric case, only gives classical
payoff. If the each player choose another strategy $\tilde U_k =
\sigma_z$, then the final state will be
\begin{eqnarray}
\tilde{\ket{\psi_f}} = \frac{1}{\sqrt2}{\hat
J}^\dagger({\sigma_x}_A\otimes {\sigma_x}_B)(\ket{00}+i\ket{11}) =
\ket{00}
\end{eqnarray}
and the expected payoff is $\{\bar\Pi_A,
\bar\Pi_B\}=\{\alpha,\alpha\}$. This value is the new Nash
equilibrium within the quantum strategy, {\it i.e.,}
$\{\bar\Pi_A(\sigma_z,\sigma_z),
\bar\Pi_B(\sigma_z,\sigma_z)\}\geq\{\bar\Pi_A(\sigma_z,\hat
U_B),\bar\Pi_B(\hat U_A,\sigma_z)\}$ for all $\hat U_k$. We can
conclude that the allowed $\cal SB$'s quantum strategies always
predict the human's choice (diagonal elements), the solution of
the game. Although the classical solution to the PD using repeated
game gives a equilibrium point in $\{\alpha, \alpha\}$~\cite{Sch}
as like as Eisert {\it et al.}, the quantum solution to the PD is
one-shot solvable. After the expected payoff is reported with the
PD, we can confirm the solution of the NP unambiguously.

\subsection{\label{sec:leve32}Quantum NP Solution}

The quantum solutions to the PD, two-person complete information
game, directly correspond to the NP, if we consider the game NP
only its reduced one-person game. The Nash equilibrium change
$\{\gamma,\gamma\}$ to $\{{\frac{1}{4}(\alpha+
\delta+\beta+\gamma), \frac{1}{4}(\alpha+\delta+\beta+\gamma)}\}$
on the Hadamard strategy. Furthermore the strategy $\sigma_z$
creates the new optimal equilibrium $\{\alpha,\alpha\}$,
corresponding to the classical {\it Pareto} optimal. Therefore we
can adapt the solution of the NP from PD, as the payoff
$\bar\Pi_k=\alpha$. Another words the human's unique choice $Box1$
is restricted by quantum mechanical $\cal {SB}$'s prophecy, maybe
$\cal SB$'s prediction is always corrected as the diagonal
elements in the given payoff matrix. This result also can be
calculated by the convex combination of the human's pure
choice~\cite{PS}, let's the human choose the $Box1$ with a certain
probability $p\in[0,1]$ and $Box2$ with a probability $1-p$, then
the intermediate state of the NP~(using the Hadamard transform
$H$) is given by
\begin{eqnarray}
\ket{\psi_f}\bra{\psi_f} = \frac{1}{2}\left(%
\begin{array}{cc}
  1 & 2p-1 \\
  2p-1 & 1 \\
\end{array}%
\right).
\end{eqnarray}\\
So that the expected payoff of the NP is
$\bar\Pi_A=\frac{1}{4}[\alpha+(2p-1)\delta+ (2p-1)\beta+\gamma]$,
coincide with the previous value in the PD. If one player choose
the quantum strategy, such as $\sigma_x$ occur a previous
situation, the Nash equilibrium $\bar\Pi_A=\alpha$, this is the
predicted value by $\cal SB$'s $\bar\Pi_B=\alpha$.

\bigskip
\section{\label{sec:leve41}Conclusion}

In this paper, we used the payoff matrix with arbitrary variables
and also described the performance of quantum strategies, which
enable to represent classical strategies. Especially we regarded
the one-person NP as a two-person PD problem and then, under
quantum strategies, obtained more reasonable and uplifted
solutions over classical strategies. In the same strategies, we
found out the Nash equilibrium always lies along the fair values,
regardless of the payoff matrix. That is, the winner is never
determined by the payoff, but by the level of strategies.
Therefore the Newcomb's paradox is not a paradox under the fair
situation.

\begin{acknowledgments}
We wish to acknowledge the support of the Korea Research
Foundation(KRF) Grant.
\end{acknowledgments}


\begin{thebibliography}{1}
\bibitem{Jeong}
Assume that the $\cal SB$'s prophecy and human's decision are
expected almost simultaneously.
\end{thebibliography}
\end{document}